# Robust inference methods of diagnostic test accuracy meta-analysis for influential outlying studies via density power divergence

Kotaro Sasaki[1,2] Hisashi Noma[1,3], Theodoros Evrenoglou[4]

[1:] The Graduate Institute for Advanced Studies, The Graduate University for Advanced Studies, Tokyo, Japan

[2:] Human Biology Creation Hub, Deep Human Biology Learning, Eisai Co. Ltd., Tokyo, Japan

[3:] Department of Interdisciplinary Statistical Mathematics, The Institute of Statistical Mathematics, Tokyo, Japan

[4:]Institute of Medical Biometry and Statistics, Faculty of Medicine and Medical Center—University of Freiburg, Freiburg im Breisgau, Germany

*Corresponding author: Kotaro Sasaki

The Graduate Institute for Advanced Studies

The Graduate University for Advanced Studies

10-3 Midori-cho, Tachikawa, Tokyo 190-8562, Japan

TEL: +81-50-5533-8514

e-mail: sasaki.kotaro.ac@gmail.com

# Abstract

In diagnostic test accuracy meta-analysis (DTA-MA), standard inference methods using bivariate random-effects models for jointly synthesizing sensitivity and specificity can be sensitive to outlying studies and may yield misleading conclusions. In this article, we propose frequentist outlier-robust statistical inference methods for DTA-MA based on density power divergence. The proposed methods automatically downweight influential outlying studies by modifying the estimating function using the robust divergence with a tuning parameter. To achieve robust yet statistically efficient inference in the presence of outlying studies, the proposed methods incorporate practical strategies for selecting the tuning parameter, including a data-adaptive criterion based on the Hyvärinen score. We also quantify the contributions of individual studies to the robust pooled estimates, facilitating interpretation of how outlying studies affect the results. We illustrate the effectiveness of the proposed methods through an application to a DTA-MA of the Mini-Mental State Examination. Simulation studies showed that the proposed methods reduced bias and root mean squared error relative to existing methods and improved coverage probability in the presence of outliers. The proposed methods enable a sensitivity analysis to assess whether the main results obtained using standard methods are driven by outlying studies.

# 1. Introduction

Meta-analysis is an essential methodology for synthesizing evidence on diagnostic test accuracy (DTA)[1, 2]. In DTA meta-analysis (DTA-MA), to account for heterogeneity among studies, diagnostic performance is commonly summarized using bivariate random-effects models, which treat sensitivity and specificity as correlated outcomes[3, 4]. However, inference based on likelihood methods in these models can be highly sensitive to influential studies whose characteristics differ markedly from the majority[5, 6]. In practice, such outlying studies may arise from differences in study design, patient spectrum, threshold effects, or data quality. Even a small number of influential studies can substantially distort pooled estimates and compromise the validity of confidence intervals (CIs), raising concerns about the robustness of standard meta-analytic conclusions.

Several effective methods have been proposed to handle outlying studies in both univariate and multivariate meta-analysis[5-11]. To the best of our knowledge, methods proposed for robust statistical inference in DTA-MA have been limited to non-normal random-effects models that accommodate outliers[5, 6]. Recently, the bivariate finite mixture (BFM) model[5] that assumes a latent mixture structure in which each study arises from either an outlying or a non-outlying group has been proposed. Another approach adopts a Laplace distribution for the random-effects[6]. A notable advantage of these models is that they can accommodate multiple outlying studies. However, existing evidence suggests that reliable statistical inference under such models may be challenging. A recent simulation study in univariate meta-analysis showed that meta-analysis models based on non-normal random-effects distributions had limited ability to provide accurate statistical inference in the presence of outliers[12], suggesting that similar difficulties may arise in DTA-MA. Indeed, simulation results for the BFM model in DTA-MA suggested that the coverage probability of its CIs was unsatisfactory[5].

In this article, we propose robust statistical inference methods for DTA-MA based on density power divergence (DPD) as an alternative approach to the outlier problem. DPD provides a generalized likelihood framework for robust inference in the presence of outliers by replacing the Kullback-Leibler divergence that underlies conventional likelihood-based inference, including maximum likelihood estimation, which is known to be sensitive to outliers[13]. A key feature of the DPD approach is that it automatically downweights influential outlying studies and allows substantial robustness to outliers with little loss of statistical efficiency through an appropriate choice of the tuning parameter[13-15]. In addition, it provides a clear framework for statistical inference grounded in estimating equation theory, and its practical utility has been demonstrated in a range of settings[11, 13, 16-18].

Although DPD-based methods have been successfully applied to pairwise meta-analysis[11], their extension to multivariate settings such as DTA-MA is nontrivial due to the need to jointly model correlated outcomes and between-study covariance structures. We address these challenges by developing DPD-based inference methods under a bivariate random-effects model. To achieve robust yet statistically efficient inference in the presence of outlying studies, the proposed methods incorporate practical strategies for selecting the tuning parameter. In addition, we introduce study-specific contribution measures that quantify the influence of individual studies on the pooled sensitivity and specificity. These measures aid interpretation of the robust pooled estimates and facilitate practical sensitivity analyses.

The rest of this article is organized as follows. In Section 2, we extend the DPD approach to DTA-MA under a bivariate random-effects model, and we present the study-specific contribution measures and tuning-parameter selection strategies. In Section 3, we illustrate the effectiveness of the proposed methods by applying them to a DTA-MA of the Mini-Mental State Examination (MMSE) for predicting progression to Alzheimer's disease dementia. In Section 4, we conduct simulation studies to evaluate the empirical performance of the proposed methods relative to existing alternatives. In Section 5, we discuss the advantages and limitations of our methods, and we provide guidance for future research.

## 2. Methods

### 2.1 Bivariate random-effects models for DTA-MA

We first describe the standard bivariate normal-normal (BNN) random-effects model for DTA-MA[3]. The BNN model synthesizes sensitivity ($Se$) and specificity ($Sp$) from $N$ independent DTA studies. Let $Y_i = \big(\text{logit}(Se_i), \text{logit}(Sp_i)\big)^T$ denote the observed logit-transformed sensitivity and specificity for study $i$ ($i = 1, 2, \ldots, N$). Under the BNN model,

$Y_i$ and $\theta_i$ are assumed to follow bivariate normal distributions:

$$Y_i \sim N_2(\theta_i, S_i),$$
$$\theta_i \sim N_2(\mu, \Sigma), \tag{1}$$

where $S_i = \begin{pmatrix} s_{i1}^2 & 0 \\ 0 & s_{i2}^2 \end{pmatrix}$ denotes the within-study covariance matrix, which is assumed to be known, and $\theta_i = (\theta_{i1}, \theta_{i2})^T$ denotes the true underlying logit-transformed sensitivity and specificity for the $i$th study. The vector $\mu = (\mu_1, \mu_2)^T$ represents the pooled logit-transformed sensitivity and specificity, and $\Sigma = \begin{pmatrix} \sigma_1^2 & \sigma_{12} \\ \sigma_{12} & \sigma_2^2 \end{pmatrix}$ represents the between-study covariance matrix. In addition, let $W_i = (S_i + \Sigma)^{-1}$ denote the inverse of the marginal covariance matrix of $Y_i$. The likelihood function of this model is given as

$$L(\mu, \Sigma) = \prod_{i=1}^{N} \frac{1}{2\pi \det(W_i^{-1})^{1/2}} \exp\left[-\frac{1}{2}(Y_i - \mu)^T W_i (Y_i - \mu)\right].$$

By maximizing this likelihood function, we obtain the maximum likelihood estimator. Instead, maximizing the restricted log-likelihood function yields the restricted maximum likelihood estimator, which is commonly used.

As an alternative, the bivariate binomial-normal (BBN) model specifies a binomial distribution for the within-study variability and a normal distribution for the between-study variability[4]. Maximum likelihood estimation can also be used under this model. However, maximum likelihood estimators derived from these models are sensitive to outlying studies with characteristics that differ markedly from those of the other studies[5].

## 2.2 The density power divergence (DPD) approach
### 2.2.1 Density power divergence
The DPD approach has been widely used in statistical analysis as an alternative framework to obtain robust estimators in the presence of outliers[13]. In general, for independent but not identically distributed observations, the DPD is defined as

$$D(\Psi; \alpha) = \frac{1}{\alpha} \sum_{i=1}^{N} f_i(Y_i; \Psi)^\alpha - \frac{1}{1+\alpha} \sum_{i=1}^{N} \int f_i(t; \Psi)^{1+\alpha} dt, \tag{2}$$

where $f_i(Y_i; \Psi)$ is the probability density function for the $i$th study, parameterized by $\Psi$,

and $\alpha$ is a tuning parameter that controls robustness to outliers $(0 < \alpha < 1)$[16]. In the DPD approach, the parameters are estimated by maximizing $D(\Psi; \alpha)$. For $\alpha > 0$, the corresponding estimating equation involves density power weights, which yield bounded influence functions for the estimators and hence robustness against outliers[16, 18]. Intuitively, DPD achieves robustness by downweighting studies with low fitted density, so potential outliers have less influence on the estimators.

When $\alpha \to 0$,

$$\lim_{\alpha \to 0}[D(\Psi; \alpha) - N\left(\frac{1}{\alpha} - 1\right)] = \log[L(\Psi)].$$

Therefore, $D(\Psi; \alpha)$ reduces to the log-likelihood as $\alpha$ approaches zero, suggesting that DPD-based estimation includes the ordinary maximum likelihood estimation as a special case. In general, increasing $\alpha$ involves a trade-off: the estimator becomes more robust to outliers, but less statistically efficient. Basu et al.[13] showed that estimators with small positive $\alpha$ can achieve substantial robustness to outliers with little loss of asymptotic efficiency relative to the maximum likelihood estimator.

### 2.2.2 DPD-based inference under the BNN model

Noma et al.[11] proposed DPD-based robust inference methods for pairwise meta-analysis. We extend this approach to DTA-MA with the BNN model. Under the BNN model (1) and the DPD defined in Equation (2), the DPD can be expressed as

$$D(\boldsymbol{\mu}, \boldsymbol{\Sigma}; \alpha) = \sum_{i=1}^{N} \left[\frac{g_i(\boldsymbol{Y}_i; \boldsymbol{\mu}, \boldsymbol{\Sigma})}{\alpha} - \frac{Q_i^\alpha}{(1+\alpha)^2}\right],$$

where

$$Q_i = (2\pi)^{-1} \det(\boldsymbol{W}_i^{-1})^{-\frac{1}{2}},$$

$$g_i(\boldsymbol{Y}_i; \boldsymbol{\mu}, \boldsymbol{\Sigma}) = Q_i^\alpha \exp\left\{-\frac{\alpha}{2}(\boldsymbol{Y}_i - \boldsymbol{\mu})^T \boldsymbol{W}_i(\boldsymbol{Y}_i - \boldsymbol{\mu})\right\}.$$

We propose the following DPD-based estimator:

$$(\widehat{\boldsymbol{\mu}}, \widehat{\boldsymbol{\Sigma}}) = \arg\max_{(\boldsymbol{\mu}, \boldsymbol{\Sigma})} D(\boldsymbol{\mu}, \boldsymbol{\Sigma}; \alpha).$$

Here, we treat the tuning parameter $\alpha$ as fixed. The choice of $\alpha$ is discussed in Section 2.2.4.

The estimating equation for $\boldsymbol{\mu}$ is given as

$$\frac{\partial D(\boldsymbol{\mu}, \boldsymbol{\Sigma}; \alpha)}{\partial \boldsymbol{\mu}} = \sum_{i=1}^{N}[g_i(\boldsymbol{Y}_i; \boldsymbol{\mu}, \boldsymbol{\Sigma})\boldsymbol{W}_i(\boldsymbol{Y}_i - \boldsymbol{\mu})] = \boldsymbol{0}. \tag{3}$$

In contrast to the estimating equation under the maximum likelihood estimation, which can be written as $\sum_{i=1}^{N}[\boldsymbol{W}_i(\boldsymbol{Y}_i - \boldsymbol{\mu})]$, $g_i(\boldsymbol{Y}_i; \boldsymbol{\mu}, \boldsymbol{\Sigma})$ acts as a weight in the DPD estimating equation. Solving Equation (3) yields

$$\boldsymbol{\mu} = \left\{\sum_{i=1}^{N} g_i(\boldsymbol{Y}_i; \boldsymbol{\mu}, \boldsymbol{\Sigma})\boldsymbol{W}_i\right\}^{-1} \sum_{i=1}^{N} g_i(\boldsymbol{Y}_i; \boldsymbol{\mu}, \boldsymbol{\Sigma})\boldsymbol{W}_i\boldsymbol{Y}_i.$$

Therefore, $\hat{\boldsymbol{\mu}}$ can be computed iteratively using a fixed-point algorithm as follows[18]:

$$\hat{\boldsymbol{\mu}}^{(k+1)} = \left\{\sum_{i=1}^{N} g_i(\boldsymbol{Y}_i; \hat{\boldsymbol{\mu}}^{(k)}, \hat{\boldsymbol{\Sigma}}^{(k)})\widehat{\boldsymbol{W}}_i^{(k)}\right\}^{-1} \sum_{i=1}^{N} g_i(\boldsymbol{Y}_i; \hat{\boldsymbol{\mu}}^{(k)}, \hat{\boldsymbol{\Sigma}}^{(k)})\widehat{\boldsymbol{W}}_i^{(k)}\boldsymbol{Y}_i, \tag{4}$$

where $\hat{\boldsymbol{\mu}}^{(k)}$, $\hat{\boldsymbol{\Sigma}}^{(k)}$, and $\widehat{\boldsymbol{W}}_i^{(k)}$ denote the estimates at the $k$th iteration of the sequential computation procedure. $\hat{\boldsymbol{\Sigma}}^{(k)}$ can be computed numerically using an iterative algorithm (e.g., the Newton-Raphson method). The proposed estimators are asymptotically normal, and empirical estimators of their asymptotic variance in sandwich form are available[16, 18].

A Wald-type CI for $\hat{\mu}_j$ is constructed as

$$\hat{\mu}_j \pm z_{\alpha/2}\sqrt{\widehat{V}_n},$$

where $z_{\alpha/2}$ denotes the upper $\alpha/2$ quantile of the standard normal distribution and $\widehat{V}_n$ denotes the variance estimate of $\mu_j$ ($j \in \{1, 2\}$). When the number of studies is small, this CI may be unreliable because it relies on an asymptotic approximation. Additionally, in such cases, outlying studies can have a particularly large influence. As an alternative expected to perform better in meta-analyses with a small number of studies, we adopt a Hartung-Knapp-Sidik-Jonkman (HKSJ)-type CI[19-21]:

$$\hat{\mu}_j \pm t_{2N-2, \alpha/2}\sqrt{\widehat{V}_n},$$

where $t_{2N-2,\alpha/2}$ is the upper $\alpha/2$ quantile of the $t$-distribution with $2N - 2$ degree of freedom.

### 2.2.3 Study-specific contribution measures

To assess how individual studies contribute to the proposed estimators of the pooled logit-transformed sensitivity and specificity, we introduce study-specific contribution measures. We define the following matrix:

$$U_i = \left\{\sum_{j=1}^{N} g_j(Y_j; \hat{\mu}, \hat{\Sigma})\widehat{W}_j\right\}^{-1} g_i(Y_i; \hat{\mu}, \hat{\Sigma})\widehat{W}_i.$$

Equation (4) implies that when the algorithm has converged and $\hat{\mu}^{(k+1)} \approx \hat{\mu}^{(k)}$, we obtain $\hat{\mu} \approx \sum_{i=1}^{N} U_i Y_i$. The matrix $U_i$ has three key properties: (1) $\sum_{i=1}^{N} U_i = I$, where $I$ denotes the identity matrix; (2) $U_i$ reduces to the standard generalized least square weight $\{\sum_{j=1}^{N} \widehat{W}_j\}^{-1}\widehat{W}_i$ as $\alpha \to 0$; and (3) for fixed $\hat{\mu}$ and $\hat{\Sigma}$, $U_i \to O$ as $\|Y_i - \hat{\mu}\| \to \infty$, indicating that $U_i$ approaches the zero matrix for outlying studies. To summarize the contributions to pooled sensitivity and specificity, we define

$$p_i^{(Se)} = \frac{|(U_i)_{11}|}{\sum_{j=1}^{N}|(U_j)_{11}|},$$

$$p_i^{(Sp)} = \frac{|(U_i)_{22}|}{\sum_{j=1}^{N}|(U_j)_{22}|}.$$

$p_i^{(Se)}$ and $p_i^{(Sp)}$ quantify how much the $i$th study's sensitivity and specificity data directly contribute to the corresponding pooled estimates, respectively. Smaller values indicate that the study is more heavily downweighted and contributes less to the pooled estimates.

In addition, $g_i(Y_i; \hat{\mu}, \hat{\Sigma})$ can be used to assess how strongly each study is downweighted in the DPD estimating equation (Equation [3]). We refer to $g_i(Y_i; \hat{\mu}, \hat{\Sigma})$ as the DPD weight. It can be interpreted as a relative weight, with 1 corresponding to the maximum likelihood estimation case. In contrast to $U_i$, the DPD weight reflects only the downweighting induced by the DPD approach and does not incorporate the precision of the data. When the results from the proposed methods differ from those obtained with standard methods, $p_i^{(Se)}$, $p_i^{(Sp)}$ and the DPD weight help identify the downweighted studies that account for the discrepancy.

**2.2.4 Choice of the tuning parameter $\alpha$**

We adopt two approaches to select the tuning parameter $\alpha$, which controls robustness to outlying studies.

The first approach uses a constant value $\alpha_{GES}$ that minimizes the gross-error sensitivity of $\boldsymbol{\mu}$, as proposed by Saraceno et al.[18]. The gross-error sensitivity is a robustness measure defined as the supremum of the absolute influence function, which quantifies the worst-case impact of a small amount of outlier contamination on an estimator[22]. The minimizer $\alpha_{GES}$ is given by $1/(p+1)$, where $p$ denotes the number of components in the study-level outcome vector $Y_i$ in the underlying statistical model. In DTA-MA synthesizing sensitivity and specificity, $p = 2$, and thus $\alpha_{GES} = 1/3$.

Alternatively, $\alpha$ can be specified in a data-adaptive manner using an efficient criterion based on the Hyvärinen score[11, 15]. The Hyvärinen score is an empirical estimate of the Fisher divergence and can be used to assess the discrepancy between a fitted unnormalized statistical model and the unknown data-generating distribution. Sugasawa and Yonekura[15] proposed selecting $\alpha$ by minimizing an asymptotic approximation of the Hyvärinen score for a pseudo-unnormalized model constructed from the DPD. In our proposed methods, the selection criterion is given as

$$H(\alpha; \widehat{\boldsymbol{\mu}}, \widehat{\boldsymbol{\Sigma}}) = \frac{1}{N} \sum_{i=1}^{N} \left[ \phi(Y_i; \widehat{\boldsymbol{\mu}}, S_i + \widehat{\boldsymbol{\Sigma}})^{\alpha} \{ \alpha (Y_i - \widehat{\boldsymbol{\mu}})^T \widehat{W}_i^2 (Y_i - \widehat{\boldsymbol{\mu}}) - tr(\widehat{W}_i) \} \right.$$
$$\left. + \frac{1}{2} \phi(Y_i; \widehat{\boldsymbol{\mu}}, S_i + \widehat{\boldsymbol{\Sigma}})^{2\alpha} (Y_i - \widehat{\boldsymbol{\mu}})^T \widehat{W}_i^2 (Y_i - \widehat{\boldsymbol{\mu}}) \right],$$

where $\phi(Y_i; \widehat{\boldsymbol{\mu}}, S_i + \widehat{\boldsymbol{\Sigma}})$ is the probability density function of $N_2(\widehat{\boldsymbol{\mu}}, S_i + \widehat{\boldsymbol{\Sigma}})$. We define $\alpha_H$ as the value of $\alpha$ that minimizes the criterion $H(\alpha; \widehat{\boldsymbol{\mu}}, \widehat{\boldsymbol{\Sigma}})$. To obtain $\alpha_H$, a grid-search algorithm can be employed: we specify a candidate set for $\alpha$ (e.g., values from 0.01 to 0.50 in increments of 0.01), compute $(\widehat{\boldsymbol{\mu}}, \widehat{\boldsymbol{\Sigma}})$ for each candidate $\alpha$, evaluate $H(\alpha; \widehat{\boldsymbol{\mu}}, \widehat{\boldsymbol{\Sigma}})$, and select the $\alpha$ that yields the minimum value.

## 3. Applications

To illustrate the effectiveness of the proposed methods, we applied them to a DTA-MA of the MMSE data reported in Arévalo-Rodríguez et al.[23]. This meta-analysis assessed the diagnostic accuracy of the MMSE for identifying individuals with mild cognitive impairment at baseline who subsequently progressed to Alzheimer's disease dementia.

**Figure 1a** presents the forest plot for this meta-analysis. Eight studies were included, with sensitivity ranging from 27% to 89% and specificity ranging from 33% to 90%. The standardized residuals were relatively large for Studies 1, 2, and 3, suggesting that these studies may be outlying and influential (**Figure 1b**).

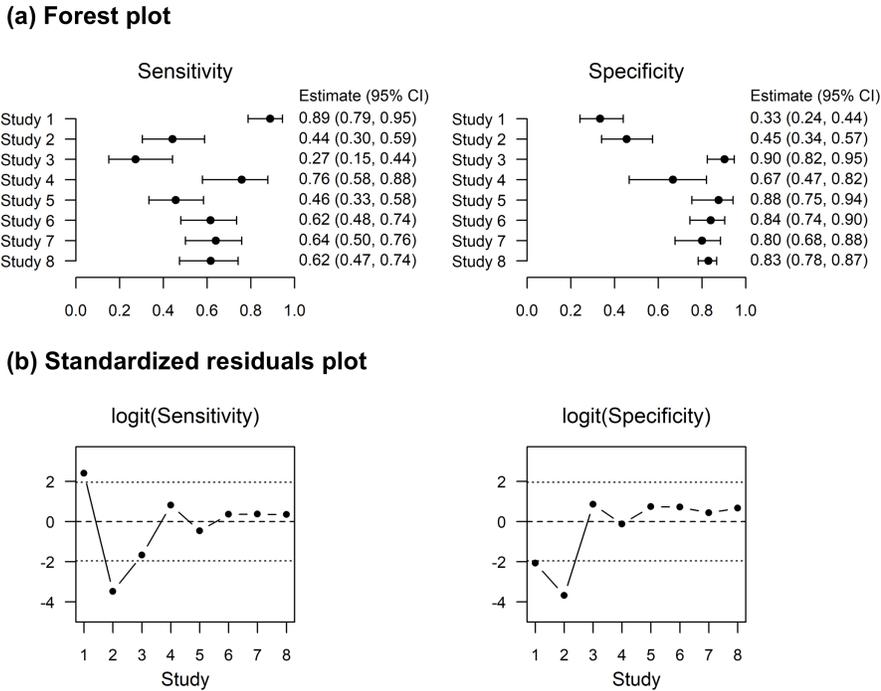

**Figure 1.** Forest plots and standardized residuals plots for the DTA meta-analysis of the MMSE data.

We applied four methods to the MMSE data: the BNN model[3], the BBN model[4], the BFM model[5], and the proposed DPD-based method under the BNN model (BNN-DPD). For the BNN-DPD, we considered two choices of the tuning parameter: $\alpha_{\text{GES}}$ and $\alpha_H$. For $\alpha_H$, we performed a grid search over candidate values from 0.01 to 0.50 in increments of 0.01. We reported both Wald-type and HKSJ-type CIs for the pooled sensitivity and specificity. The DTA-MA dataset and R codes for implementing the proposed robust inference methods are available on GitHub: https://github.com/KSasaki7/Robust_DTAMA_DPD.

**Table 1** summarizes the results of the four methods. The BNN and BBN models yielded almost identical estimates of the pooled sensitivity and specificity. The BFM model provided similar point estimates, but slightly narrower CIs. In contrast, the BNN-DPD yielded pooled specificity estimates 7 to 9 percentage points higher than the BNN model, whereas sensitivity estimates were essentially unchanged. The CIs of the proposed methods tended to be wider, particularly when using $\alpha_{\text{GES}}$. The HKSJ-type CIs were slightly wider than the corresponding Wald-type CIs. In addition, the Hyvärinen-score criterion selected value of $\alpha_H = 0.50$, suggesting that a DPD-based model that downweights outlying studies was preferred over the maximum likelihood estimation case ($\alpha_H \approx 0$).

**Table 1.** Results of the DTA meta-analysis of the MMSE data.

| Method | Sensitivity (95% CI) | (HKSJ-type 95% CI) | Specificity (95% CI) | (HKSJ-type 95% CI) | $\alpha$ |
|---|---|---|---|---|---|
| BNN | 0.60 (0.45–0.74) | — | 0.74 (0.58–0.86) | — | — |
| BBN | 0.60 (0.45–0.73) | — | 0.74 (0.59–0.85) | — | — |
| BFM | 0.60 (0.48–0.72) | — | 0.75 (0.63–0.84) | — | — |
| BNN-DPD ($\alpha_{\text{GES}}$) | 0.61 (0.28–0.86) | (0.26–0.88) | 0.81 (0.42–0.96) | (0.38–0.97) | 0.33 |
| BNN-DPD ($\alpha_H$) | 0.59 (0.41–0.75) | (0.39–0.76) | 0.83 (0.78–0.88) | (0.77–0.88) | 0.50 |

**Abbreviations:** BNN, bivariate normal-normal model; BBN, bivariate binomial-normal model; BFM, bivariate finite mixture model; BNN-DPD, bivariate normal-normal model with density power divergence; CI, confidence interval; HKSJ, Hartung-Knapp-Sidik-Jonkman.

**Figure 2** presents the DPD weights and the contribution rates of individual studies for the pooled sensitivity and specificity under the BNN-DPD using $\alpha_{\text{GES}}$. Overall, Study 2, which reported both sensitivity and specificity below 50%, had the lowest DPD weight and contribution rate, followed by Studies 1 and 3, indicating that these studies were substantially downweighted by the proposed method. The higher pooled specificity can be explained by downweighting Studies 1 and 2, which reported low specificity. In contrast, the sensitivity values in these downweighted studies were heterogeneous, which may explain why the DPD approach had little impact on the pooled sensitivity estimate.

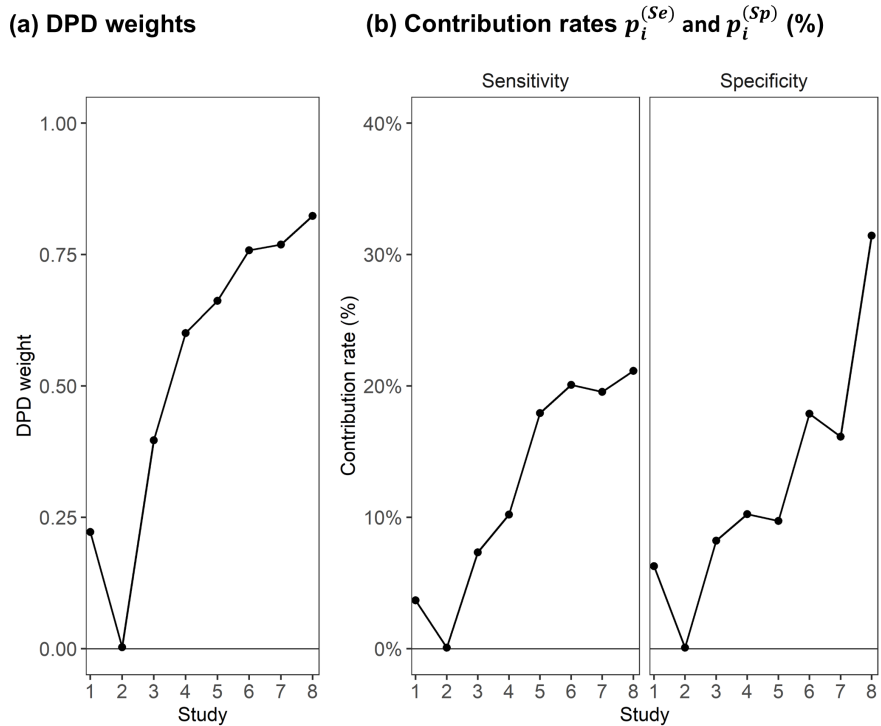

**Figure 2.** The DPD weights and the contribution rates for the pooled sensitivity and specificity of individual studies under the proposed method using $\alpha_{\text{GES}}$.

## 4. Simulations

In this section, we describe the methods for the simulation studies conducted to evaluate the performance of our proposed methods, using the ADEMP framework[24], and then present the results.

### 4.1 Aim and data generating mechanisms

The aim of the simulation studies was to evaluate the performance of the proposed methods in comparison with existing methods, under conditions with and without outlying studies.

Simulation data were generated under the BBN model, motivated by the DTA-MA of the MMSE data in Arévalo-Rodríguez et al.[23] described in Section 3. Study-level true sensitivities and specificities were first sampled from a bivariate normal distribution on the logit scale, and two-by-two tables were then generated by binomial sampling conditional on study-specific numbers of diseased and non-diseased participants. For each study, the sample size was randomly drawn from a discrete uniform distribution between 53 and 351. The disease prevalence was generated from a continuous uniform distribution between 13% and 53%.

We further assumed that each study belonged to either the outlying-study group or the non-outlying-study group. For the non-outlying studies, $\boldsymbol{\mu}$ was set to the logit-transformed values corresponding to $(Se, Sp) = (0.61, 0.82)$. In contrast, for the outlying studies, either sensitivity or specificity was set to a lower value, namely $(0.27, 0.82)$ or $(0.61, 0.39)$. Note that studies in which both sensitivity and specificity are below 50%, as in Study 2 in the MMSE data, are rare in practice and were therefore not considered in our simulations. The between-study heterogeneity parameters were set to $(\tau_1^2, \tau_2^2, \rho) = (0.15, 0.10, -0.7)$ and were assumed to be common to both the outlying and non-outlying studies.

**Table 2** summarizes the simulation settings. A total of 15 settings were considered, defined by the number of studies and the configurations of the outlying studies. The number of studies $N$ was set to 8, 12, and 16. For the configurations of the outlying studies, we considered five scenarios (A)–(E): $(k_1, k_2) = (0,0), (2,0), (0,2), (2,1), (1,2)$, where $k_1$ and $k_2$ are the numbers of studies with outlying sensitivity $(Se = 0.27)$ and outlying specificity $(Sp = 0.39)$, respectively. Scenario (A) includes no outlying studies $(k_1 = k_2 = 0)$. For each scenario, we generated 1,000 datasets.

**Table 2.** Summary of the simulation settings.

| Number of studies | $N = 8, 12, 16$ |
|---|---|
| **Scenarios for outlying studies** | |
| (A) | No outlying studies: $(k_1, k_2) = (0, 0)$ |
| (B) | Two studies with outlying sensitivity: $(k_1, k_2) = (2, 0)$ |
| (C) | Two studies with outlying specificity: $(k_1, k_2) = (0, 2)$ |
| (D) | Two studies with outlying sensitivity and one with outlying specificity: $(k_1, k_2) = (2, 1)$ |
| (E) | One study with outlying sensitivity and two with outlying specificity: $(k_1, k_2) = (1, 2)$ |

### 4.2 Estimands, Methods, and Performance Measures

The estimands of interest were the pooled sensitivity and specificity among the non-outlying studies, set to $(Se, Sp) = (0.61, 0.82)$.

We compared four methods: the BNN model, the BBN model, the BFM model, and the proposed methods (BNN-DPD). As in Section 3, for the proposed methods, we considered two approaches for tuning parameter selection and reported both Wald-type and HKSJ-type CIs. For $\alpha_H$, we performed a grid search over candidate values from 0.01 to 0.50 in increments of 0.01.

The performance measures were absolute bias and root mean squared error (RMSE) for the pooled sensitivity and specificity among the non-outlying studies, as well as the coverage probability and the width of the 95% CIs. All performance measures were computed using only replications for which the algorithms converged.

### 4.3 Simulation results

We present the simulation results in **Figures 3–4**. In terms of bias and RMSE, the proposed methods and the existing methods performed comparably under Scenario (A), which included no outlying studies. Under Scenarios (B)–(E), which involved outlying studies in sensitivity and/or specificity, the existing methods, including the BFM model, exhibited larger bias. In contrast, the proposed BNN-DPD methods were consistently less biased and achieved lower RMSE. Regarding tuning parameter selection, $\alpha_{GES}$ generally yielded slightly better performance than $\alpha_H$. Across all methods, bias tended to be larger when the number of studies was small.

Regarding the 95% CIs, coverage probability tended to fall below the nominal level across all methods. Under Scenario (A), the proposed method using $\alpha_{GES}$ with HKSJ-type intervals achieved the best coverage in most cases, whereas the proposed method using $\alpha_H$ showed poorer coverage. Under Scenarios (B)–(E), the CIs of the existing methods had coverage below the nominal level in most cases, particularly for the BFM model. In contrast, the proposed method using $\alpha_{GES}$ produced coverage closer to the nominal level; with HKSJ-type intervals, coverage for both sensitivity and specificity was consistently above 90%.

Compared with $\alpha_{GES}$, the proposed method using $\alpha_H$ yielded narrower intervals but also lower coverage. In most cases, a larger number of studies led to higher coverage and narrower intervals. An exception was the BFM model, for which specificity coverage deteriorated in Scenarios (C) and (E).

Overall, convergence failures were infrequent across the methods and simulation settings. The proposed method using $\alpha_H$ failed to converge in only one replicate in one setting, whereas the BFM model showed 1 to 4 non-converged replicates in five settings. All other methods converged in all settings.

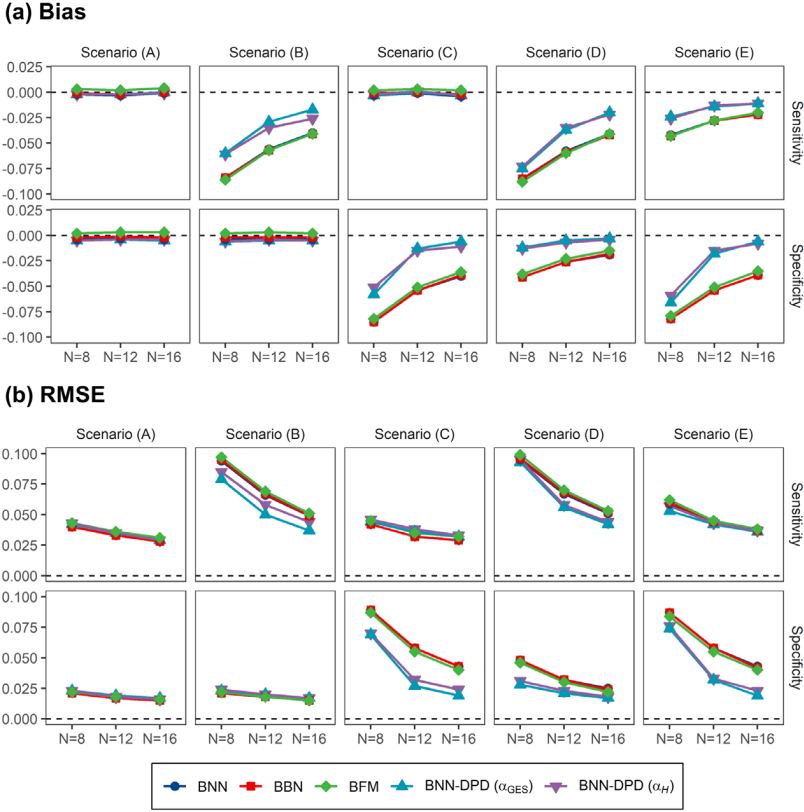

**Figure 3.** Simulation results: bias and RMSE for pooled sensitivity and specificity estimates. The five scenarios (A)–(E) correspond to settings of $(k_1, k_2) = (0,0), (2,0), (0,2), (2,1), (1,2)$, where $k_1$ and $k_2$ are the numbers of studies with outlying sensitivity and outlying specificity, respectively. (BNN, bivariate normal-normal model; BBN, bivariate binomial-normal model; BFM, bivariate finite mixture model; BNN-DPD, bivariate normal-normal model with density power divergence)

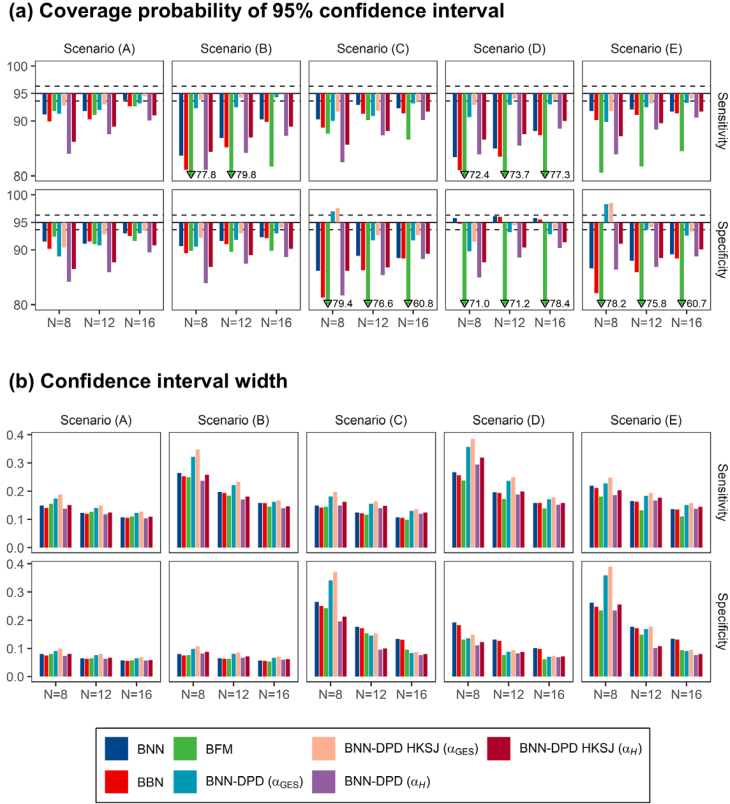

**Figure 4.** Simulation results: coverage probability and mean width of 95% confidence intervals. The five scenarios (A)–(E) correspond to settings of $(k_1, k_2) = (0,0), (2,0), (0,2), (2,1), (1,2)$, where $k_1$ and $k_2$ are the numbers of studies with outlying sensitivity and outlying specificity, respectively. In panel (a), triangles indicate bars that were truncated for visual clarity, and the corresponding actual values are provided alongside. The dashed lines represent the upper and lower bounds of the 95% confidence interval for the nominal level based on 1,000 simulation replicates. (BNN, bivariate normal-normal model; BBN, bivariate binomial-normal model; BFM, bivariate finite mixture model; BNN-DPD, bivariate normal-normal model with density power divergence, BNN-DPD HKSJ, BNN-DPD with Hartung-Knapp-Sidik-Jonkman-type confidence intervals.)

## 5. Discussion

In DTA-MA, the standard estimators based on bivariate random-effects models can be sensitive to influential outlying studies and may lead to misleading conclusions. We proposed robust inference methods based on DPD that accommodate such outlying studies. The proposed methods provide robust pooled estimates and confidence intervals while automatically downweighting influential outlying studies. We discussed two practical strategies for selecting the robustness tuning parameter $\alpha$: a fixed choice based on gross-error sensitivity, $\alpha_{\text{GES}}$, and a data-adaptive choice, $\alpha_H$. We also provided study-specific
16

contribution measures to aid interpretation. Overall, the proposed methods can serve as a robust complement to standard bivariate random-effects meta-analysis, helping assess the sensitivity of conclusions to outlying studies.

In the application example, only the proposed methods mitigated the impact of potential outlying studies and yielded a much higher pooled specificity estimate than the existing methods. This finding suggests that existing methods alone may not always be sufficient to account for the influence of outlying studies in practice. In the simulation studies, the existing methods, including the BNN and BBN models and the outlier-accommodating BFM model, exhibited larger bias when outlying studies were present. Consistent with previous findings, the BFM model showed much lower coverage probabilities under outlier contamination. Compared with these methods, the proposed methods yielded consistently better performance in terms of bias and RMSE, regardless of whether $\alpha_{GES}$ or $\alpha_H$ was used. For CIs, using $\alpha_{GES}$ with HKSJ-type CIs generally achieved the best coverage probabilities, although the resulting intervals were wider. Because $\alpha_H$ is selected by minimizing a score computed from estimated model components, the selection may be unstable and not always yield an appropriate value, particularly when the number of studies is small. Therefore, we recommend using the proposed method with $\alpha_{GES}$ and HKSJ-type CIs when conducting analyses that account for outlying studies. This strategy would mitigate the impact of outlying studies and provide more reliable inference.

One useful feature of the proposed methods is that the study-specific contribution measures quantify how strongly each study influences the pooled estimates, thereby facilitating interpretation of the results. In practice, researchers should explore study characteristics that may explain the discrepancy among studies, especially in heavily downweighted studies. Although the standardized residual and other influence diagnostic methods[9, 10] can also be used to quantify the influence of outlying studies, their ability may be limited when multiple outliers are present simultaneously[25]. In addition, compared with the existing methods based on non-normal random-effects models, the proposed methods are often computationally less demanding. Moreover, their firm grounding in estimating equation theory provides a principled framework, which in turn makes the methods a reliable choice for practical use.

One potential limitation of the proposed methods is that their performance may be suboptimal when the number of studies is small, as indicated by the simulation results. The CIs tended to undercover, although HKSJ-type intervals partially mitigated this issue. Another possible approach to improve the CI coverage is to adopt bootstrap methods for estimators based on estimating equations[26, 27]. Extending the proposed methods to provide more reliable inference with a limited number of studies is an important subject for future research. In addition, the proposed methods can be extended to other multivariate meta-analysis settings, including network meta-analysis, where outlying studies may lead to problematic inconsistency between direct and indirect evidence [28, 29]. Although the proposed methods were developed under the BNN model, the DPD approach is applicable to other models in DTA-MA as well. However, for models that do not admit a closed-form expression for the marginal likelihood, such as the BBN model, the DPD cannot be derived analytically, and additional computational strategies would be required[30]. These extensions also represent important directions for future research.

In conclusion, we proposed robust inference methods for DTA-MA that are effective in the presence of outlying studies. Using the proposed methods, researchers can perform a sensitivity analysis to assess whether the main results obtained using standard methods are sensitive to outlying studies.

## Acknowledgements

KS and HN were funded by Grants-in-Aid for Scientific Research from the Japan Society for the Promotion of Science (grant numbers: JP23K11931, JP22H03554, and JP24K21306). TE was funded by the Deutsche Forschungsgemeinschaft (DFG, German Research Foundation) under Project ID 554095932.

## Declaration of Conflicting Interest

KS is an employee of Eisai Co., Ltd., and HN received a research fund from GlaxoSmithKline and consulting fees from GlaxoSmithKline, Sony, and Kowa, outside the submitted work.

**Data Availability Statement**

The diagnostic test accuracy meta-analysis dataset used in Section 3 and the R code are available on GitHub: https://github.com/KSasaki7/Robust_DTAMA_DPD.